\documentclass[
aps, prb,twocolumn,floatfix,
amsmath,amssymb,superscriptaddress,bibnotes,longbibliography
]{revtex4-2}

\usepackage[utf8]{inputenc}
\usepackage{hyperref}
\usepackage{graphicx}
\usepackage{mathtools}
\usepackage{color}
\usepackage{xcolor}
\usepackage{diagbox}
\usepackage{tabularx}
\newcolumntype{Y}{>{\centering\arraybackslash}X}

\usepackage{tabu}
\usepackage{verbatim}
\usepackage{amsmath}
\usepackage{stmaryrd}

\usepackage{tikz}

\usepackage[export]{adjustbox}
\usepackage{comment}
\usepackage{amssymb}

\begin{document}

\title{Inchworm quasi Monte Carlo for quantum impurities}

\author{Hugo U.~R.\ Strand}
\email{hugo.strand@oru.se}
\affiliation{School of Science and Technology, Örebro University, SE-701 82 Örebro, Sweden}
\affiliation{Institute for Molecules and Materials, Radboud University, 6525 AJ Nijmegen, the Netherlands}
\author{Joseph Kleinhenz}
\affiliation{Lawrence Berkeley National Laboratory, University of California, Berkeley, CA 94720-8229, USA}
\author{Igor Krivenko}
\affiliation{Institut für Theoretische Physik, Universität Hamburg, Notkestraße 9, 22607 Hamburg, Germany}

\date{\today}

\begin{abstract}
  The inchworm expansion is a promising approach to solving strongly correlated quantum impurity models due to its reduction of the sign problem in real and imaginary time.
  However, inchworm Monte Carlo is computationally expensive, converging as $1/\sqrt{N}$ where $N$ is the number of samples.
  We show that the imaginary time integration is amenable to quasi Monte Carlo, with enhanced $1/N$ convergence,
  by mapping the Sobol low-discrepancy sequence from the hypercube to the simplex with the so-called Root transform.
  This extends the applicability of the inchworm method to,
  e.g., multi-orbital Anderson impurity models with off-diagonal hybridization,
  relevant for materials simulation,
  where continuous time hybridization expansion Monte Carlo has a severe sign problem.
\end{abstract}

\maketitle
\makeatletter
\let\toc@pre\relax
\let\toc@post\relax
\makeatother

Quantum impurity models are key for describing the physics of low dimensional quantum systems such as quantum dots \cite{hansonSpinsFewelectronQuantum2007}, molecular junctions \cite{scottKondoResonancesMolecular2010}, and cold atom systems \cite{blochManybodyPhysicsUltracold2008} which can exhibit quantum many-body effects like the Coulomb blockade \cite{beenakkerTheoryCoulombblockadeOscillations1991} and Kondo screening \cite{kondoResistanceMinimumDilute1964, wilsonRenormalizationGroupCritical1975}.
Quantum impurity models are also central for the ab-initio simulation of correlated quantum materials where they appear as auxiliary problems within dynamical mean-field theory \cite{Georges:1996aa} and its extensions \cite{RevModPhys.90.025003, Ayral:2015ab, PhysRevB.94.075159, Sun:2002aa}.
Therefore, efficient simulation of quantum impurity models is of paramount importance to improving our understanding of quantum dots, correlated materials and their emergent collective phenomena, like anomalous superconductivity \cite{kancharlaAnomalousSuperconductivityIts2008}, metal--insulator transitions \cite{rozenbergMottHubbardTransitionInfinite1994}, and magnetic order \cite{shinaokaPhaseDiagramPyrochlore2015, zhangMetalInsulatorTransitionTopological2017}.

In thermal equilibrium, continuous time quantum Monte Carlo \cite{Prokofev:1998fu, PhysRevB.72.035122, Gull:2011lr} provides numerically exact solutions to a wide range of quantum impurity problems, e.g.\ multi channel quantum dots connected to separate reservoirs and correlated materials with high symmetry.
For multi-channel and multi-orbital impurities, the continuous time hybridization expansion algorithm (CTHYB) \cite{Werner:2006rt, Werner:2006qy, Haule:2007ys} has been particularly important for progress in the field.
However, for intermixed reservoirs, low crystal symmetry, and materials with relativistic mixing of spin and orbital momentum, the majority of continuous time quantum Monte Carlo algorithms suffer from a sign problem \cite{Gull:2011lr}.

One notable exception is inchworm quantum Monte Carlo \cite{Inchworm,InchwormEquil}, where causality is exploited to reorder the diagram summation.
The reordering has been shown to give sub-exponential decay of the average sign in real time \cite{Inchworm}, as well as a drastic improvement in the sign for low symmetry equilibrium problems, where CTHYB suffers from a dire sign problem \cite{InchwormEquil}.
However, the attainable accuracy is limited by the $1/\sqrt{N}$ convergence of the integration with the number of Monte Carlo steps $N$.
Additionally, the order-by-order Monte Carlo integrals are normalized by overlaps with low order diagrams, which becomes difficult at lower temperatures \cite{InchwormGF}.

\newpage

\begin{figure}[b]
\ \\[-5mm]
\includegraphics[scale=1.0] {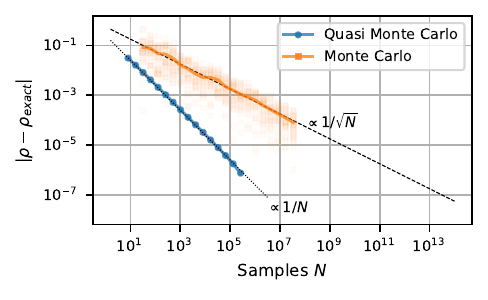}\ \\[-6mm]
\caption{Absolute error $|\rho - \rho_{exact}|$ in the many body density matrix $\rho$ as a function of the number of samples $N$ for the proposed inchworm quasi Monte Carlo algorithm (blue circles) with $1/N$ scaling (dotted line) compared to the Monte Carlo based continuous time hybridization expansion \cite{Werner:2006rt, Werner:2006qy, Haule:2007ys, Parcollet2015398, Seth2016274} (orange squares) with $1/\sqrt{N}$ scaling (dashed line). \label{fig:qqmc_scaling}}
\end{figure}

In this letter, we show that by using quasi Monte Carlo integration \cite{dick_kuo_sloan_2013, MCQMC2016, Nuyens+2014+223+256} in imaginary time the convergence rate can be improved from $1/\sqrt{N}$ to $1/N$, greatly increasing the efficiency of the algorithm (see Fig.~\ref{fig:qqmc_scaling}).
Additionally, since quasi Monte Carlo is a direct integration technique there is no need for order-by-order normalization.
We benchmark our quasi Monte Carlo inchworm algorithm on several impurity models computing both the local many-body density matrix as well as single- and two-particle response functions.
%

\begin{figure}
\includegraphics[scale=1.0] {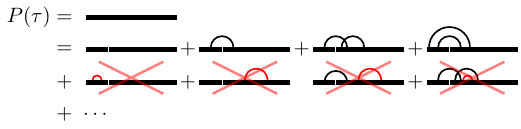}\\[-4mm]
\caption{Modified inchworm expansion for the pseudo-particle propagator $P(\tau)$ with bold propagators on both sides of $\tau_w$. \label{fig:diags}}\ \\[-7mm]
\end{figure}

\textit{Inchworm algorithm.}
%
In its original formulation, the inchworm quantum Monte Carlo method is a stochastic strong coupling technique for solving general quantum impurity models \cite{Inchworm}.
The algorithm relies on the causality property of the pseudo-particle propagator $P(\tau)$ \cite{Eckstein:2010fk}, which is computed step by step starting at $\tau_i = 0$ and ending at $\tau_f = \beta$ where $\beta$ is the inverse temperature.
At each step, a short (compared to $\beta$) new segment $[\tau_w; \tau_f]$ is attached to the previously computed segment $[0; \tau_w]$, see Fig.\ \ref{fig:diags}.
Once $ P(\tau)$ has been determined, the single-particle Green's functions $G_{\alpha\beta}(\tau) = - \langle \mathcal{T} c_\alpha(\tau) c_\beta^\dagger(0)\rangle$ is computed as a sum over another set of diagrams \cite{InchwormGF}.
%

\begin{figure}[t]
\includegraphics[scale=0.5] {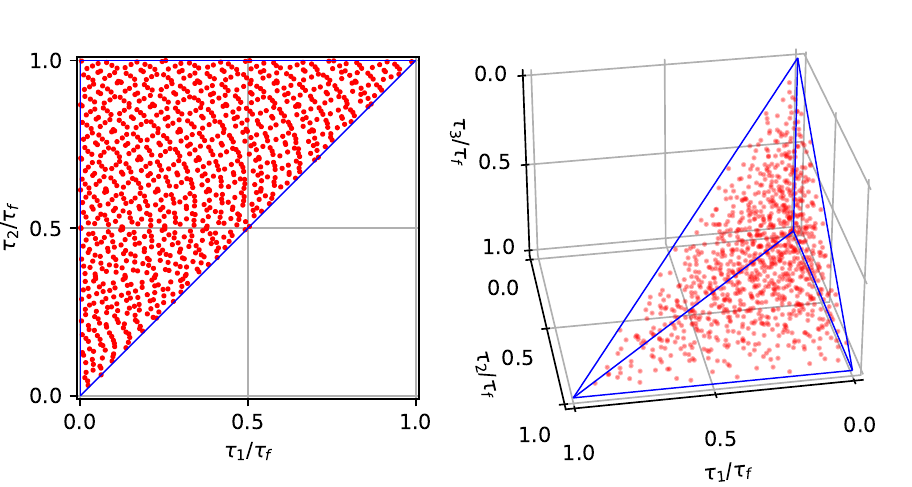}\\[-4mm]
\caption{Sobol sequences with 1024 points mapped from the hypercube to the simplex in two (left panel) and three (right panel) dimensions using the Root transform \cite{PILLARDS200529} in Eq.\ (\ref{eq:root}). \label{fig:sobol_simplex}}\ \\[-7mm]
\end{figure}

\textit{Quasi Monte Carlo integration on the simplex.}
%
For each step of the inchworm algorithm all possible \emph{inchworm proper} \cite{Inchworm, InchwormGF} diagrams have to be integrated over the internal imaginary times. The internal times $\tau_j$ are ordered, i.e.\ $\tau_f \geq \tau_1 \geq \tau_2 \geq \ldots \geq \tau_d \geq 0$, and the integration domain is a $d$-dimensional simplex.
However, the low-discrepancy sequences used for quasi Monte Carlo integration, like the Sobol sequence \cite{Sobol1966} used in our calculations, are commonly defined in the $d$-dimensional hypercube $\vec{x} \in [0, 1]^d$.
Hence, a transformation that maps the $d$-dimensional hypercube to the $d$-dimensional simplex, while preserving the low discrepancy property of the transformed sequence is required.
We adopt the Root transformation \cite{PILLARDS200529}\\[-4mm]
\begin{equation}
  \tau_1 = \tau_f x_1^{1/d}
  , \quad
  \tau_2 = \tau_1 x_2^{1/(d-1)}
  , \, \ldots , \quad
  \tau_d = \tau_{d-1} x_d
  \, ,
  \label{eq:root}
\end{equation}
which is continuous, has a constant Jacobian $\tau_f^d / d!$, and with linear complexity in the dimension $d$ is computationally feasible in high dimensions.
Figure \ref{fig:sobol_simplex} shows the distribution of points in the two- and three-dimensional $\tau$-simplex obtained by Root-transforming the first 1024 points of the Sobol sequence.

Unlike an earlier proposed transformation based on an exponential model function \cite{PhysRevLett.125.047702}, the Root transformation has no free adjustable parameters and does not discard any points from the input low-discrepancy sequence. The latter feature has recently been shown to be of crucial importance \cite{SobolDigitalNets}, as skipping even one point spoils the digital net property of the sequence and can worsen the convergence rate.
Furthermore, numerical experiments with individual low order inchworm diagrams showed that it outperforms the alternatives from \cite{PILLARDS200529, PhysRevLett.125.047702} in terms of the relative error scaling with the number of taken samples.

\textit{Inchworm Quasi Monte Carlo.}
%
Our implementation of the inchworm algorithm does not rely on any form of importance sampling.
Instead, the imaginary time integrals are computed using quasi Monte Carlo integration \cite{dick_kuo_sloan_2013, MCQMC2016, Nuyens+2014+223+256}. The remaining discrete parameters labeling various contributions to the strong coupling series -- such as expansion order, spin projections and orbital indices -- are explicitly summed over.
This approach is straightforward, well controlled yet fast enough for the moderate expansion orders $n_{max}$ explored here ($n_{max}\lesssim 7$).

At a given inchworm step defined by ($\tau_w, \tau_f$), the accumulation of the bold pseudo-particle propagators proceeds as follows.
For each expansion order $n$ up to a fixed $n_\mathrm{max}$, a list of $n!$ diagram topologies is generated.
A topology is a partition of the tuple $\{1,\ldots,2n\}$ into pairs. Diagrammatically, is can be thought of as a set of $n$ undirected hybridization lines connecting pairs of vertices located at positions $\{1,\ldots,2n\}$ on the segment $[0; \tau_f]$.
If each hybridization line is connected by crossings to a line that overarches $\tau_w$, the corresponding topology is \emph{inchworm proper} \cite{Inchworm, InchwormGF}, c.f.\ Fig.\ \ref{fig:diags}.
Note, this differs slightly from previous definitions because we choose to use the \emph{bold} rather than bare propagator between $\tau_w$ and $\tau_f$. This adjustment marginally reduces the number of diagrams appearing in the expansion.
All the proper topologies are then transformed into diagrams by replacing the paired vertices with all possible pairs of operators $c_\alpha$ and $c_\beta^\dagger$ (swapping $c$ and $c^\dagger$ within a pair results in a different diagram).
Finally, a sum of all contributing diagrams of order $n$ is integrated over the imaginary time domain by evaluating it at the first $N=2^m$ Root-transformed points of a $2n$-dimensional Sobol sequence. The power-of-two sample size restriction to $N$ is essential to preserve the digital net property of the Sobol sequence; Using other sample sizes, e.g.\ $10^m$, serverely degrades accuracy \cite{SobolDigitalNets}.

As is known in the context of hybridization expansion methods \cite{Haule:2007ys}, a significant fraction of diagrams stemming from a retained topology do not actually contribute due to symmetry considerations. We employ the autopartition algorithm \cite{TRIQSCTHYB} to reveal invariant subspaces of the atomic Hamiltonian and to unitarily transform the matrix-valued bare pseudo-particle propagator $P_0(\tau)$ into a block-diagonal form. The choice of the invariant subspaces accounts for any potential symmetry breaking caused by a non-diagonal hybridization function. This way, we ensure that the subspace partition is not too refined, and that boldified propagator $P(\tau)$ possesses the same block-diagonal form.

A Julia package implementing the proposed method is publicly available at \cite{QInchwormGithub}. It relies on the \texttt{Keldysh.jl} framework \cite{KeldyshGithub} for handling pseudo-particle propagators and Green's functions, and on the \texttt{KeldyshED.jl} package \cite{KeldyshEDGithub} for solving the atomic problem and construction of the bare atomic propagators $P_0(\tau)$.

\textit{Benchmarks.}
%
\begin{figure}[t]
\includegraphics[scale=1.0] {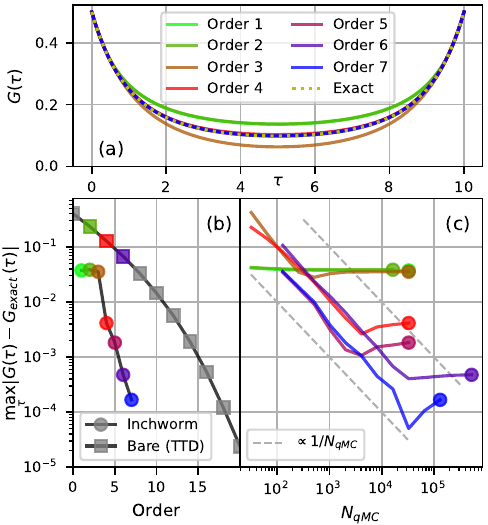}\ \\[-3mm]
\caption{Panel~(a): Imaginary time single-particle Green's function $G(\tau)$ from inchworm quasi Monte Carlo at expansion orders $n_{max}= 1$ to 7 (solid lines) compared to the exact solution (dashed line) at $\beta=10$. Panel~(b): Expansion order convergence of the error $\max_\tau | G(\tau) - G_{exact}(\tau)|$ (circles) compared to the bare TTD expansion error (at $\tau = \beta/2$) (squares) \cite{PhysRevB.107.245135}. Panel~(c): Convergence with respect to quasi Monte Carlo samples $N_{\text{qMC}}$ at fixed order. \label{fig:gf_scaling}}\ \\[-8mm]
\end{figure}
%
The theory of quasi Monte Carlo integration \cite{dick_kuo_sloan_2013, MCQMC2016, Nuyens+2014+223+256} only promises $1/N$ convergence for well behaved integrands \footnote{Strictly speaking the rate can be proven to scale as  $N^{-1}\log(N)^d$ where $d$ is the dimension for sufficiently smooth integrands.} and a crossover to the generic Monte Carlo scaling $1/\sqrt{N}$ for ill-behaved integrands. Hence, it is important to perform a direct investigation of the convergence rate when applying quasi Monte Carlo to a new class of problems.

To establish the convergence rate we study an impurity problem with a known analytical solution, namely a single fermionic state hybridized with a semi-infinite chain with nearest neighbor hopping. Since this system is non-interacting it can be solved exactly and it is also a worst case scenario impurity model for the strong coupling expansion, far from the atomic limit.
The action $\mathcal{S}$ of the impurity model has the form\\[-5mm]
\begin{equation}
  \mathcal{S}[c, c^\dagger] =
  -\!\!\int_0^\beta \!\!\! d\tau d\tau'\, 
  c^\dagger(\tau) \left[-\delta(\tau \!-\! \tau') \partial_\tau - \Delta(\tau \!-\! \tau') \right] c(\tau')
  \, ,
  \label{eq:action}
\end{equation}
where $c^\dagger$ and $c$ are creation and annihilation operators of the spinless fermion, and $\Delta(\tau)$ is the hybridization function of the semi-infinite chain. It is derived from a semi-circular density of states  with bandwidth $W$ coupled to the impurity with the hybridization strength $V$.

The convergence rate of the many-body density matrix $\rho = P(\beta)$ as a function of number of samples $N$ per inchworm step, in Fig.\ \ref{fig:qqmc_scaling}, is monotonic with a clear $1/N$ scaling. The difference compared to standard partition function Monte Carlo and its $1/\sqrt{N}$ convergence is clear. The improved $1/N$ rate of convergence of the inchworm quasi Monte Carlo is key to reach high precision results.
Here $\beta = 8$, $V=1/4$, and a hybridization function with bandwidth $W=2$ shifted by 1 (with the lower band edge at the Fermi level) was used, in order to stay away from half-filling, where
the inchworm quasi Monte Carlo many-body density matrix immediately converges.

Computing dynamical response functions is central for the application of an impurity solver in many-body methods such as dynamical mean-field theory \cite{Georges:1996aa}.
Therefore, we compute the single particle Green's function $G(\tau) = -\langle \mathcal{T} c(\tau) c^\dagger(0) \rangle$ \cite{InchwormGF} and study the convergence for the action in Eq.\ (\ref{eq:action}) with $W = 4$ and $V = 1$.
To see the effect of the expansion order we perform calculations including all diagrams up to a given order $n_{max}$ and observe rapid convergence of $G(\tau)$ with respect to $n_{max}$, see Fig.\ \ref{fig:gf_scaling}b. Already at order four the inchworm result for $G(\tau)$ in Fig.\ \ref{fig:gf_scaling}a is indistinguishable from the exact result.
Looking in detail at the order by order convergence in Fig.\ \ref{fig:gf_scaling}b, we observe rapid exponential convergence setting in at order 4. For a fixed order, the inchworm expansion (circles) is markedly more accurate than the bare expansion (squares) recently combined with tensor train decomposition (TTD) \cite{PhysRevB.107.245135}.
We also study the convergence rate of $G(\tau)$ with respect to the number of quasi Monte Carlo integration points. For fixed order we generically observe an initial $1/N$ rate of convergence followed by a plateau as the error becomes dominated by the difference between the fixed order result and the exact infinite order solution, which at e.g.\ order 7 is of the order $10^{-4}$, see Fig. \ref{fig:gf_scaling}c.

Note that the difference in order-by-order convergence in Fig.\ \ref{fig:gf_scaling}b is not related to the quasi Monte Carlo (and the TTD of Ref.\ \onlinecite{PhysRevB.107.245135}) but rather coming from the difference between the inchworm and bare strong coupling perturbation expansions. In terms of amount of diagrams per order, the bare expansion only contains diagrams up to a given order while the inchworm expansion contains infinite resummations of bare diagrams.
In the benchmark we see that the inchworm expansion with more diagrams converges faster with respect to order. However, this comes at the cost of the factorial scaling in the number of diagrams, while the bare expansion has a Wick's theorem and can use a determinant approach to rid of the factorial scaling \cite{Werner:2006rt}. Note, however, that the inchworm factorial scaling can be reduced to exponential using summation based on the inclusion-exclusion principle \cite{PhysRevB.98.115152}.

\begin{figure}
\includegraphics[scale=1.0] {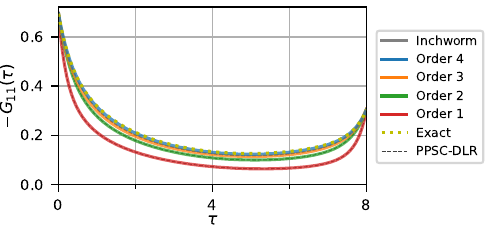}\ \\[-3mm]
\caption{
  Single particle Green's function $G_{11}(\tau)$ for the two-band Kanamori model [Eq.\ (\ref{eq:kanamori})] with off-diagonal hybridization at $\beta = 8$. The inchworm quasi Monte Carlo result (solid lines) is compared to the exact result from exact diagonalization (dashed yellow line) and the 1st, 2nd, and 3rd order result of bold pseudo-particle self-consistent perturbation theory using the discrete Lehmann representation (PPSC-DLR) \cite{2023arXiv230708566K} (thin dashed lines). \label{fig:two_band}}
\ \\[-9mm]
\end{figure}

Having established the $1/N$ convergence we now turn to benchmark our inchworm quasi Monte Carlo solver for a non-trivial multi-orbital impurity model relevant for real-materials simulation within dynamical mean-field theory.
To this end we study a spin-ful two-band impurity model with local Kanamori interaction \cite{Kanamori:1963aa, Georges:2013fk}\\[-7mm]
\begin{multline}
  H =
  \sum_{i=1}^2 U n_{\uparrow,i} n_{\downarrow,i}
  +
  \sum_{\sigma,\sigma'} (U - (3-\delta_{\sigma,\sigma'}) J) n_{\sigma,1} n_{\sigma,2}
  \\[-2mm] -
  J (
  c^\dagger_{\uparrow, 1} c^\dagger_{\downarrow, 1} c_{\uparrow, 2} c_{\downarrow, 2}
  +
  c^\dagger_{\uparrow, 1} c^\dagger_{\downarrow, 2} c_{\uparrow, 2} c_{\downarrow, 1}
  + \textrm{h.c.})
  \label{eq:kanamori}
\end{multline}
\ \\[-5mm]
where $c^{(\dagger)}_{\sigma,i}$ creates (annihilates) a fermion in state $i=1,2$ with spin $\sigma=\uparrow,\downarrow$, 
with Hubbard interaction $U=2$ and Hund's coupling $J=1/5$, chemical potential $\mu = (3U - 5J - 3)/2$, and a hybridization function,  $\Delta_{i \sigma, j \sigma'}(\tau) = \delta_{\sigma, \sigma'} \Delta(\tau)$, where $\Delta(i\nu_n) = 1/(i\nu_n + \epsilon) + 1/(i\nu_n - \epsilon)$ and $\epsilon = 2.3$.
This class of impurity models is out of reach for standard partition function continuous time hybridization expansion Monte Carlo, since the off-diagonal hybridization produces a severe sign problem \cite{Gull:2011lr}. Previously this model has been used to showcase that stochastic inchworm Monte Carlo is not impacted by such a sign problem \cite{InchwormEquil}.
Using inchworm \emph{quasi} Monte Carlo we observe a rapid convergence of the solution with expansion order, with order 4 already being remarkably close to the exact solution, obtainable by exact diagonalization since the hybridization function is a dicrete set of poles, see Fig.\ \ref{fig:two_band}.
We also compare the first, second, and third order inchworm result with bold pseudo-particle perturbation theory recently obtained from a direct integration approach \cite{2023arXiv230708566K} using the discrete Lehmann representation \cite{PhysRevB.105.235115, KAYE2022108458}. The agreement confirms that the two expansions are equivalent at fixed perturbation order.
%

\begin{figure}[t]
\includegraphics[scale=1.0] {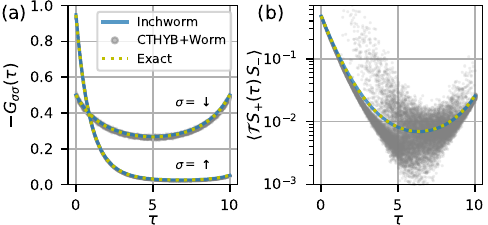}\ \\[-3mm]
\caption{Panel (a): Single particle Green's function for the spin-polarized Anderson impurity model at inverse temperature $\beta=10$. Panel (b): Out-of-plane spin-response $\langle \mathcal{T} S_{+}(\tau) S_{-}\rangle = \langle \mathcal{T} (c^\dagger_\uparrow c_\downarrow)(\tau) (c^\dagger_\downarrow c_\uparrow) \rangle$.
  The inchworm result (blue lines) at order 5 is shown together with
  the exact analytic solution (dotted yellow lines),
  and
  the continuous time hybridization expansion worm sampling (CTHYB+Worm) \cite{PhysRevB.92.155102, PhysRevB.94.125153, WALLERBERGER2019388} (grey circles) where the ergodicity problem induces an extremely large variance between bins adjacent in time.
  \label{fig:SpSm}
}
\ \\[-6mm]
\end{figure}

The inchworm approach is not limited to computing single particle response functions and is able to compute arbitrary dynamical response functions \cite{InchwormGF}. As an example we study the spin-polarized Anderson impurity model with
action
$\mathcal{S}_B = \sum_{\sigma=\uparrow,\downarrow} \mathcal{S}[c_\sigma, c^\dagger_\sigma] - \int_0^\beta \!\! d\tau \, B c^\dagger_\uparrow(\tau) c_\uparrow(\tau)$
where the field $B=2$ couples to the spin-up component and $\mathcal{S}[c_\sigma, c^\dagger_\sigma]$ (see Eq.\ \ref{eq:action}) has a spin-diagonal hybridization function $\Delta_\sigma(\tau) = \Delta(\tau)$ with semi-circular density of states, bandwidth $W=2$, and hybridization $V=1/2$.
Since the model is non-interacting, it amounts to a worst case scenario for strong coupling expansions like the inchworm approach, which expands in the hybridization function.
The single particle Green's function $G_{\sigma,\sigma}(\tau) = -\langle \mathcal{T} c_\sigma(\tau) c^\dagger_{\sigma} \rangle$ is shown in Fig.\ \ref{fig:SpSm}a with $G_{\downarrow\downarrow}$ half-filled and $G_{\uparrow\uparrow}$ almost unoccupied. The inchworm expression for dynamical response functions is agnostic to the kind of operators \cite{InchwormGF}, so computing the transverse spin-response $\langle \mathcal{T} S_+(\tau) S_- \rangle$ is straightforward, see Fig.\ \ref{fig:SpSm}b.
On the other hand, this correlator can not be sampled in partition function continuous time hybridization expansion Monte Carlo since in this case $S_+$ and $S_-$ do not commute with the local Hamiltonian \cite{Parcollet2015398, Seth2016274}. This limitation can, in part, be overcome by worm sampling \cite{PhysRevB.92.155102, PhysRevB.94.125153, WALLERBERGER2019388}. However, for some higher order correlators like $\langle \mathcal{T} S_+(\tau) S_- \rangle$ also worm sampling has ergodicity problems, as shown in Fig.\ \ref{fig:SpSm}b.
%

\textit{Discussion and conclusions.}
%
In conclusion, we have demonstrated a quantum impurity solver based on quasi Monte Carlo integration techniques \cite{dick_kuo_sloan_2013, MCQMC2016, Nuyens+2014+223+256} applied to the inchworm method \cite{Inchworm, InchwormEquil, InchwormGF}.
Using quasi Monte Carlo, a $1/N$ convergence rate with the number of samples is achievable, which compares favorably to the $1/\sqrt{N}$ convergence of previous Monte Carlo methods.
In order to achieve this convergence rate, the smooth Root transformation \cite{PILLARDS200529} of the Sobol sequence from the unit hypercube to the simplex is used, which maintains the digital net property.
We demonstrate the improved convergence of our method on simple single orbital problems where exact results are available, as well as on a more challenging multi-orbital problem.
These results show the usefulness of quasi Monte Carlo for achieving high accuracy of imaginary time integrals at moderate orders arising in diagrammatic expansions for quantum systems.

We also point out that stochastic inchworm quantum Monte Carlo previously has been applied to non-equilibrium real-time dynamics of quantum impurity problems, with considerable success in taming the dynamical sign problem \cite{Inchworm, InchwormGF}. Hence, the extension of the quasi Monte Carlo integration approach to out-of-equilibrium real-time simulations is an interesting venue for further research.

\textit{Acknowledgments.}
%
HURS and IK acknowledge funding from the European Research Council (ERC) under the European Union’s Horizon 2020 research and innovation programme (Grant agreement No.\ 854843-FASTCORR).
The computations were enabled by resources provided by the National Academic Infrastructure for Supercomputing in Sweden (NAISS) and the Swedish National Infrastructure for Computing (SNIC)
through the projects
SNIC 2022/1-18, 
SNIC 2022/6-113, 
SNIC 2022/13-9, 
SNIC 2022/21-15, 
NAISS 2023/1-44, 
and 
NAISS 2023/6-129 
at PDC, NSC and CSC partially funded by the Swedish Research Council through grant agreements no.\ 2022-06725 and no.\ 2018-05973.

\clearpage

\bibliography{manuscript}

\end{document}